\DeclareUnicodeCharacter{2062}{} 
\documentclass[reprint,a4paper,superscriptaddress,amsmath,amssymb,aps,pra,11pt,floatfix,tightenlines]{revtex4-2}

\usepackage{graphicx}
\usepackage{xcolor}
\usepackage{dcolumn}
\usepackage{bm}
\usepackage{floatrow} 
\usepackage[all]{nowidow}
\usepackage{siunitx}[=v2]
\usepackage{hyperref}
\usepackage{ulem}
\newcommand{\pr}{$\varphi_{\omega-2\omega}$}
\newcommand{\tr}{$\theta_{\omega-2\omega}$}

\begin{document}

\title{
Probing broken time-reversal symmetry with tailored-light photocurrents
}

\author{Daniel M. B. Lesko}
\affiliation{Department of Physics, Friedrich-Alexander-Universität Erlangen-Nürnberg (FAU), Erlangen, Germany}
\affiliation{Fakult\"at f\"ur Physik, Ludwig-Maximilians-Universit\"at, M\"unchen, Germany}	
\author{Tobias Weitz}
\affiliation{Department of Physics, Friedrich-Alexander-Universität Erlangen-Nürnberg (FAU), Erlangen, Germany}
\affiliation{Fakult\"at f\"ur Physik, Ludwig-Maximilians-Universit\"at, M\"unchen, Germany}	
\author{Simon Wittigschlager}
\affiliation{Department of Physics, Friedrich-Alexander-Universität Erlangen-Nürnberg (FAU), Erlangen, Germany}
\author{Selina Nöcker}
\affiliation{Department of Physics, Friedrich-Alexander-Universität Erlangen-Nürnberg (FAU), Erlangen, Germany}
\affiliation{Fakult\"at f\"ur Physik, Ludwig-Maximilians-Universit\"at, M\"unchen, Germany}
\author{Weizhe Li}
\affiliation{Department of Physics, Friedrich-Alexander-Universität Erlangen-Nürnberg (FAU), Erlangen, Germany}
\affiliation{Fakult\"at f\"ur Physik, Ludwig-Maximilians-Universit\"at, M\"unchen, Germany}
\author{Peter Hommelhoff}
\affiliation{Department of Physics, Friedrich-Alexander-Universität Erlangen-Nürnberg (FAU), Erlangen, Germany}	
\affiliation{Fakult\"at f\"ur Physik, Ludwig-Maximilians-Universit\"at, M\"unchen, Germany}	
\author{Ofer Neufeld}
\affiliation{Technion Israel Institute of Technology, Faculty of Chemistry, Haifa 3200003, Israel}

\date{\today}

\maketitle

\onecolumngrid
\vspace{-\baselineskip}
\textbf{
Light-field-driven photocurrents represent a powerful tool for generating photocurrents without external bias in light-matter systems that lack inversion symmetry.
While these photocurrents are used in electronic applications, such as current sources,
switches, and photovoltaics, their presence can also be used to probe material properties in and out of equilibrium, such as topology. 
Here we advance this path of light-field-driven photocurrent spectroscopy by utilizing tailored laser fields for ultrafast photocurrent generation to study time-reversal symmetry (TRS) broken phases.
We employ combinations of bichromatic linearly-polarized laser beams that individually respect mirror (spatial) and time-reversal symmetry, individually precluding photocurrents, but when combined can break symmetries and generate photocurrents. 
We show, both theoretically and experimentally, that unique choices of the relative polarization angle and two-color phase imposes a forbidden photocurrent selection rule in TRS-invariant systems, as the tailored light maintains TRS while breaking all other spatial symmetries. 
We then employ state-of-the-art \textit{ab-initio} simulations to validate this physical mechanism, and, crucially, predict its breaking in materials with intrinsically-broken TRS, creating a background free signal for magnetism and Chern physics. Our work paves way for probing TRS-broken phases of matter in an ultrafast time-resolved manner, not requiring the application of external magnetic fields or even circularly-polarized electric fields. 
}
\twocolumngrid

\maketitle

\vspace{\baselineskip}
In recent years ultrafast and nonlinear optics has become an incredibly useful tool for probing the unique properties of quantum materials both in and out of equilibrium. 
For example, high harmonic spectroscopy of solids~\cite{Ghimire2018,Goulielmakis2022,Heide2024} has shed light on ultrafast exciton dynamics~\cite{Heide2022,Freudenstein2022,ChangLee2024,Molinero2024,Jensen2024}, superconductivity~\cite{Alcal2022}, strongly-correlated materials~\cite{Bionta2021,Uchida2022}, topology~\cite{Bai2020,Schmid2021,Lv2021,Heide2022B,Neufeld2023B,UzanNarovlansky2024}, as well as nonlinear phononics~\cite{Bionta2021,Neufeld2022,Zhang2024}. 
Besides high harmonic spectroscopy, other nonlinear optical observables can also be employed to probe material properties, such photocurrents generated from the bulk photogalvanic effect (BPGE)~\cite{Ma2023,Pettine2023}. 
This effect allows generating photocurrents without external bias in light-matter systems that lack inversion symmetry.
Importantly, BPGE originating in inversion symmetric materials must result from the broken inversion and time reversal symmetry of the applied fields.
More generally, bulk photogalvanic photocurrents result from shift and/or injection currents, the former originating from a shift in the center of mass of conduction band population (due to the inversion asymmetry of a material) while the latter is from asymmetric conduction band population throughout the Brillouin zone (and can occur in inversion symmetric materials).

Bulk photogalvanic photocurrents provide a sensitive probe to both the material composition, as well as the optical field, resulting in a sensitive tool for investigating the combined symmetries of the light-matter system.
Interestingly, these currents have been predicted to be quantized in Weyl semimetals~\cite{Chan2017,deJuan2017} as well as give access to dynamical evolution of Floquet topological insulators (FTIs)~\cite{McIver2019,lesko2025} and quantum materials~\cite{Ju2017,Orenstein2021,Han2021,Yang2022}, hinting towards their inherent sensitivity to the underlying symmetries.

One particularly interesting realization of nonlinear optical spectroscopies employs laser driving with tailored-fields, which permits breaking or inducing symmetries beyond those intrinsic to the crystal lattice~\cite{Habibovi2024}. 
Bichromatic optical fields represent a well known and experimentally straightforward example of these tailored electric fields; extensively used in both gas and solid state high harmonic generation~\cite{Watanabe1994,Eichmann1995,Andiel1999,Dudovich2006,Fleischer2014,Hickstein2015}, have provided enhanced control of steady-state Floquet phases~\cite{Trevisan2022,Wang2023,Zhu2024,Neufeld2025}, and give access to sub-optical-cycle changes to the band structure~\cite{UzanNarovlansky2022,Mitra2024,Tyulnev2024}. 
Furthermore, bichromatic light can be used to generate and coherently control photocurrents in centrosymmetric solids~\cite{Atanasov1996,Hach1997,Sun2010,Rioux2011,Sternemann2013,Sederberg2020b,Bharti2024}, valley selectivity~\cite{JimenezGalan2020,Mrudul2021,Tyulnev2024}, and ultrafast magnetic impulses~\cite{Yuan2015,Sederberg2020}. 
However, so far, these methods were not employed to probe ultrafast magnetism~\cite{Walowski2016,Siegrist2019,Neufeld2023} or magnetic phases of matter. 

Most commonly, magnetic phenomena are measured by utilizing a probe that breaks break time-reversal symmetry (TRS) itself, accomplished with external magnetic fields or monochromatic circularly-polarized optical electric fields. 
These methods give rise to  circular-dichroic responses that are material-specific responses to the light's helicity, e.g. in transient absorption~\cite{Mak2012,Yang2015}, perturbative nonlinear optics~\cite{Herrmann2025}, photoemission spectroscopies~\cite{Wang2013,Boschini2024}, or high harmonics~\cite{Heinrich2021}.
However, so far it was unclear how one can probe phases of matter with broken TRS without employing probes (such as a circularly polarized field) that directly break TRS themselves. Beyond being a fundamental open research question, such methodologies are essential for ultrafast spectroscopies where probe pulses might affect the dynamics being measured (e.g. generate ultrafast magnetism instead of probing existing magnetism).

Here we connect the research fields of tailored light and broken time-reversal symmetry spectroscopy. 
We develop and experimentally demonstrate a spectroscopic technique utilizing BPGE photocurrents driven by femtosecond tailored-light in two-dimensional (2D) materials to understand broken time-reversal symmetry processes and materials. 
Key to our analysis lies in the addition of the $2\omega$ field, which allows for us to alter the characteristic symmetries and chirality of the probe~\cite{Neufeld2025}, providing new insights into symmetry breaking phenomena that are incapable with monochromatic fields. 

To gain an intuitive understanding of our tailored light photocurrent spectroscopy technique, we benchmark our new method by measuring photocurrents in graphene (a high symmetry and low dimensional material) driven by polarization-tailored $\omega-2\omega$ bichromatic pulses.
Graphene is an inversion and time-reversal symmetric material, resulting in BPGE currents from primarily momentum space population asymmetries, i.e, injection currents, present after the passage of the pulse until relaxation of the conduction band population.
Unique to our tailored light photocurrent spectroscopy technique, each individual carrier component of the driving beam is linearly-polarized and hence does not generate a BPGE in graphene or break TRS on its own.
By tuning the relative angle between the two linearly polarized waves, as well as their relative phase, the bichromatic beam can either (i) break all point-group symmetries of the crystal (including TRS) and generate a BPGE current; or (ii) break all point-group symmetries besides TRS, leading to a TRS-based photocurrent suppression selection rule.
By measuring the angle and phase-dependent photocurrent we map the symmetry elements of the combined light-matter interaction in a precise and controllable manner.
Importantly, the latter bichromatic field can probe TRS without inherently breaking it, starkly different to what has been done traditionally with circular monochromatic fields.
We experimentally show these selection rules for graphene, and further validate this physical mechanism with \textit{ab-initio} time-dependent density functional theory (TDDFT) calculations, and an analytical theory. 
We then employ \textit{ab-initio} simulations in the 2D inversion symmetric hexagonal magnet, CrI$_{3}$~\cite{Huang2017}, showing that the photocurrent selection rule breaks due to the broken TRS intrinsic to the material, leading to a unique system-specific behavior. 
Finally, we find that this selection rule breaks in simulations of a Floquet topological insulator (FTI, a Chern insulator with broken TRS~\cite{Oka2009,Lindner2011}) generated by irradiation of graphene with a circularly-polarized dressing mid-infrared laser.
Our theoretical analysis, and experimental benchmark, allows us to establish that BPGE driven by bichromatic fields provides a background-free signal for TRS-broken phases that can be used for ultrafast spectroscopy of Chern (quantized) physics and magnetism. 

\vspace{\baselineskip}
\noindent \textbf{Concept}
 
\begin{figure*}[h!]
	\includegraphics[width=\linewidth]{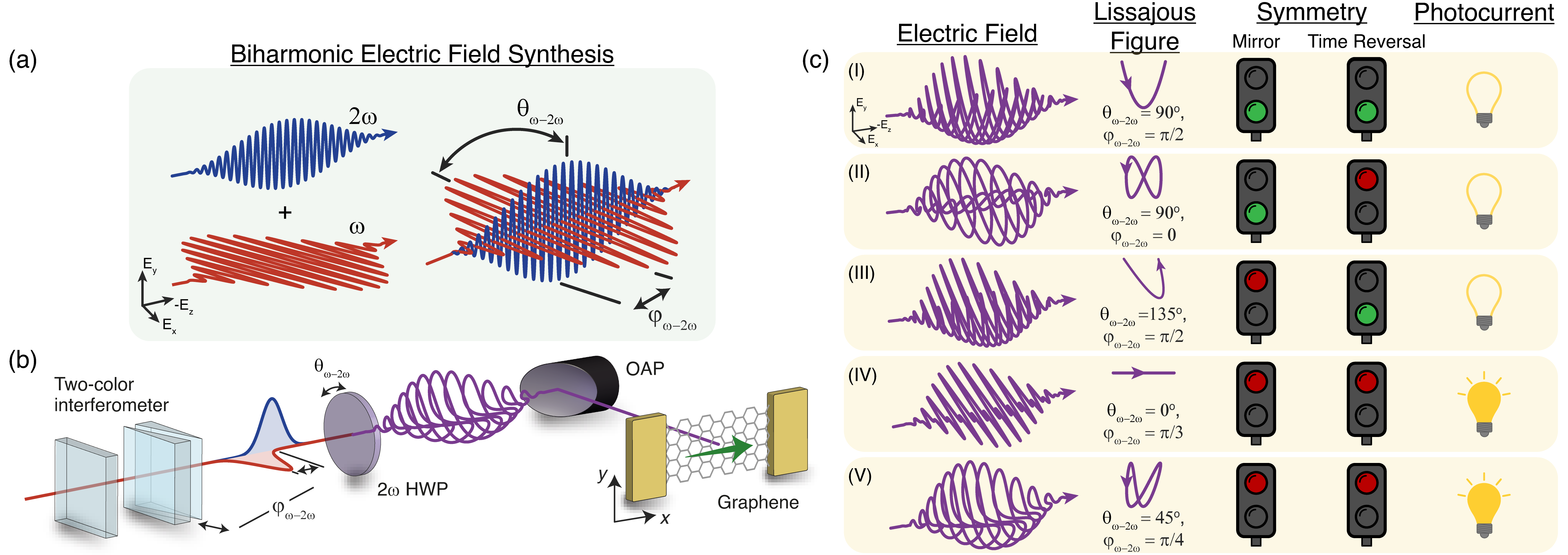}
    \caption{\textbf{Tailored light two-color symmetry breaking.} \textbf{a}, Linearly polarized optical harmonics are combined with control over the relative angle ($\theta_{\omega-2\omega}$) and the two-color delay ($\varphi_{\omega-2\omega}$). \textbf{b}, Sketch of the experimental setup. The phase ($\varphi_{\omega-2\omega}$) between the co-propagating $\omega$ and $2\omega$ pulses is controlled in a collinear two-color interferometer. With a dichroic half-waveplate ($2\omega$ HWP), we control the angle of the $2\omega$ field with the $\omega$ polarization fixed (\tr). Using an off-axis parabolic mirror (OAP), we focus the resulting waveform on a monolayer graphene strip, mounted in a vacuum chamber (not shown). We measure $\theta_{\omega-2\omega}$ and $\varphi_{\omega-2\omega}$-dependent photocurrents via gold electrodes attached to the graphene strip. \textbf{c}, When combined, these fields can break mirror symmetry (in the x direction) and/or time-reversal symmetry, with broken symmetry indicated by the red traffic light, depending on both $\theta_{\omega-2\omega}$ and $\varphi_{\omega-2\omega}$, generating a photocurrent in graphene as indicated by the yellow light-bulb. We showcase exemplary broken symmetry waveforms in cases I-V.}
    \label{Fig1}
\end{figure*}

We begin by outlining the concept of the employed biharmonic $\omega-2\omega$ tailored-light (Fig.~\ref{Fig1}a).
This biharmonic field has two key degrees of freedom: The relative polarization angle between the two beams (\tr), and the relative phase of the two carrier components, \pr. 
Crucially, we can tune \tr\ and \pr\ by adjusting the angle of a two-color half-wave plate and the thickness of a calcite plate, respectively, in a two-color inline interferometer (Fig.~\ref{Fig1}b)~\cite{Brida2012}. This way, we effectively alter the characteristic shape and Lissajous of the resulting two-color driving field (Fig.~\ref{Fig1}c), and measure \pr-dependent photocurrents arising along the \textit{x}-axis (Fig.~\ref{Fig1}b).
Further experimental details on the setup and the biharmonic field synthesis are delegated to the Methods section. 

These degrees of freedom in the light field allow us to explore two main symmetry elements in the light-matter interaction: a mirror symmetry along the \textit{y(-z)}-plane (MS, labeled $\sigma_y$, invoking $x \rightarrow -x$) and time-reversal symmetry (TRS, invoking $\mathrm{\textbf{E}}(t) \rightarrow \mathrm{\textbf{E}}(-t)$). 
For instance, for \tr$=90^\circ$ and \pr$=\pi/2$ (Fig.~\ref{Fig1}c I) the optical electric field exhibits a distinct dynamical mirror symmetry~\cite{Neufeld2019B} such that we have the symmetry relation $\mathrm{\textbf{E}}(t)=\sigma_y \cdot \mathrm{\textbf{E}}(t+T/2)$, where $\sigma_y$ is coupled to time-translations of half an optical period ($T=2\pi/\omega$).
When this field interacts with graphene, the full light-matter Hamiltonian respects all symmetry elements because graphene is itself both mirror and time-translation symmetric. This symmetry in the Hamiltonian leads to the photocurrent generation along the \textit{x}-direction being forbidden (Fig.~\ref{Fig1}c I).

Interestingly, the separate control of the angle and phase of the beam components importantly allows for complete and independent mirror and TRS breaking (see Fig.~\ref{Fig1}c II-V), which cannot be achieved solely with monochromatic optical fields. That is, by proper choice of the laser parameters we can either invoke or break each of these symmetry elements separately, allowing us to probe the material symmetries without having the probe pulse break the symmetry being investigated.
This allows utilizing the biharmonic synthesis to explore TRS breaking phenomena in a unique manner - without the presence of any other symmetry elements in the system such as mirror or inversion.
This is a crucial point in our analysis - by allowing selective degrees of freedom to be scanned in a multi-dimensional manner, we can develop a spectroscopy technique that is sensitive to independent symmetry breaking elements, since we can probe the system's characteristic response as the field transitions in between the symmetric and symmetry-broken states. 
This principle is similar to symmetry-breaking spectroscopy techniques previously proposed for chiral phenomena~\cite{Fischer2002,Ayuso2019,Neufeld2019,Mayer2024}, but utilizes nonlinear photocurrent generation that removes the need for atomic strength optical fields.
Importantly, using photocurrent suppression to investigate symmetries should be less sensitive to material decoherence pathways (e.g. electron-electron and electron-phonon scattering) since the current signal accumulates, and similarly should be insensitive to phase-matching that affects optical schemes.

\vspace{\baselineskip}
\noindent \textbf{Results}

\begin{figure*}[h!]
	\includegraphics[width=\linewidth]{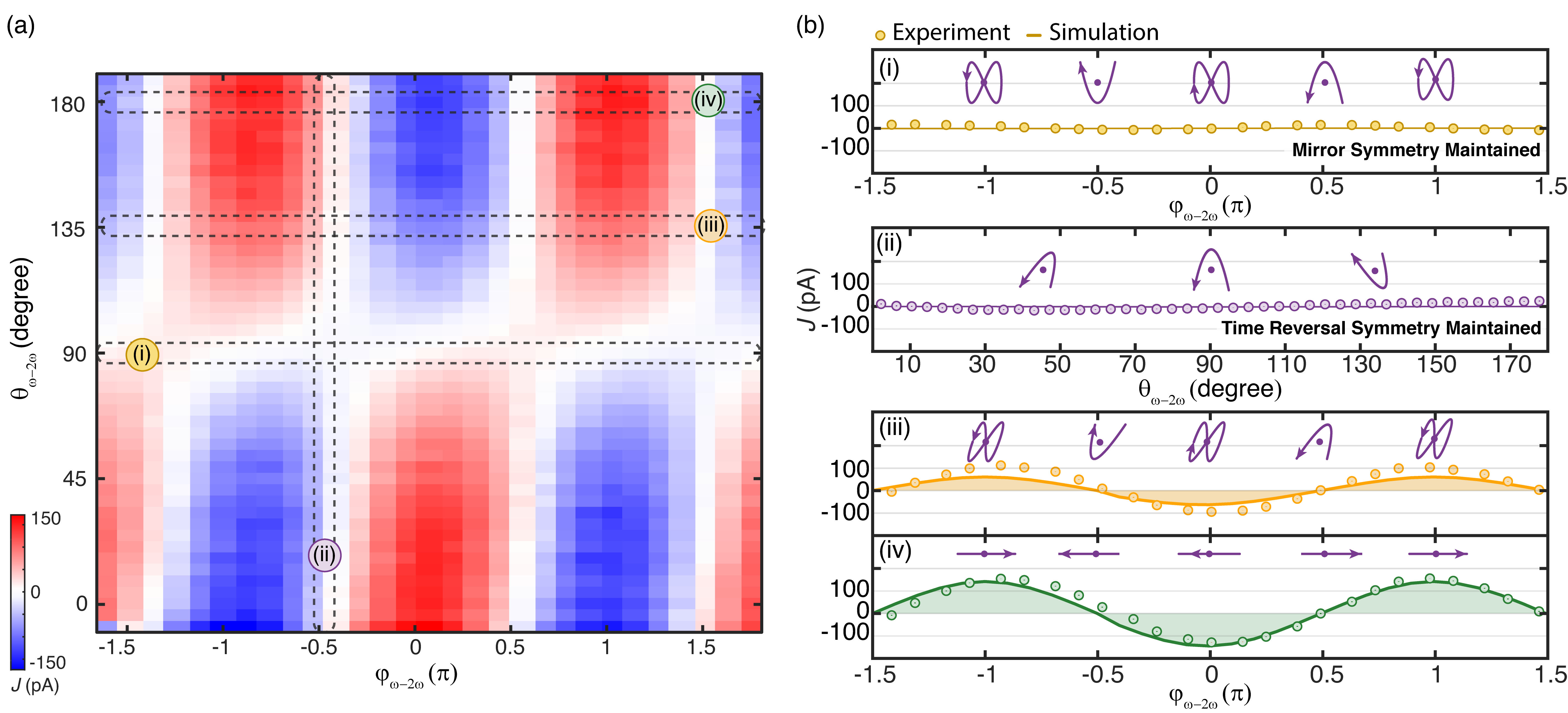}
    \caption{\textbf{Two-color photocurrent spectroscopy.} \textbf{a}, Two-color phase ($\varphi_{\omega-2\omega}$) and angle ($\theta_{\omega-2\omega}$)-dependent photocurrents in graphene. Regions of zero photocurrent correspond to waveforms that exhibit mirror and/or time-reversal symmetry. \textbf{b (i-iv)}, linecuts of (a) showcasing different symmetries.
    \textbf{b (i)} A horizontal line-cut at \tr$=90^\circ$ exhibits photocurrent suppression for all \pr\ due to the presence of mirror planes in the material.
    \textbf{b (ii)} A vertical line-cut at \pr$=\pi/2$ showcasing photocurrent suppression for all \tr\ due to the material and waveform maintaining TRS for all phases.
    \textbf{b (iii)} Linecut at \tr$=135^\circ$: both mirror and TRS are broken independently, allowing for \pr-dependent photocurrent generation.
    \textbf{b (iv)}, Finally \pr-dependent photocurrents, at \tr$=180^\circ$, are generated from the simultaneous breaking of both mirror and TRS.}
    \label{Fig2}
\end{figure*}

By measuring photocurrents along the \textit{x}-direction as a function of the two degrees of freedom (\tr\ and \pr), we now map out the symmetry-dependent photocurrent selection rules of graphene. Our main experimental data is presented in Fig~\ref{Fig2}, showing measured BPGE photocurrents. The currents exhibit a characteristic structure that reflects the combined symmetries of the material and optical electric field. To further understand these complex features, we analyze specific cases (Fig.~\ref{Fig2}b i-iv) for the \tr\ and \pr-dependent photocurrent.
Below, we will show that all regions of Fig~\ref{Fig2}a exhibiting a photocurrent suppression result from either (or both) TRS and MS being maintained in the light-matter system.

First, for \tr$=90^\circ$ (Fig.~\ref{Fig2}b i), we observe a photocurrent suppression for all phases \pr .
Specifically, this example of a photocurrent suppression is a fundamental consequence of a more generalized dynamical \textit{mirror symmetry} within the driving field, as discussed above, which was previously analyzed and observed in refs.~\cite{Neufeld2021,lesko2025}.
This feature can be indicative of the presence of mirror planes in graphene, since if the selection rule is upheld, it follows that the material also exhibits the appropriate mirror plane. However, it is not helpful for deciphering if the system respects or breaks TRS.

Another robust feature in the photocurrent map (Fig.~2a) are the minima for \pr$=\pi/2 \pm m\pi$ (for integer $m$) and all \tr . Here the tailored field is clearly polarized in the two-dimensional space of the monolayer plane (the \textit{xy}-plane), for every choice of $\theta$ (Fig.~\ref{Fig2}b ii).
In this case, mirror symmetry is broken by the optical field for any generic choice of \tr$\neq0,180^\circ$ (see Lissajous figures inset in Fig.~\ref{Fig2}b ii), implying a different symmetry is inducing this selection rule. 
Indeed, it is apparent that these parameters invoke TRS for all \tr\ (i.e. $\mathbf{E}(t)=\mathbf{E}(-t)$ regardless of the choice of \tr). Therefore, the vanishing photocurrent is a result of a \textit{TRS-induced selection rule}, in absence of any other symmetry relation that might cause such suppression. 
In that respect, by measuring photocurrents as a function of \pr\ across regions where the field's symmetry is tuned, one can study the light-matter system as it transitions from a non-symmetric to symmetric state via photocurrent spectroscopy. 
Vanishing photocurrents for the correct \tr\ and \pr\ (\tr$\neq90^\circ$, \pr$=\pi/2$, for example Fig.~\ref{Fig1}c III) indicate a symmetry-relation intrinsic to the material system, which can be probed by the photocurrent amplitudes. Further, the magnitude and location of the photocurrent nodes (i.e. their potential deviation from the expected selection rule) can be used as indicators for the extent of an intrinsic broken symmetry.
Fig.~\ref{Fig2}b iii and iv shows such a region where MS and TRS are broken and revive, allowing us to explore their individual roles in photocurrent generation.
Finally, while the co-polarized case (Fig.~\ref{Fig2}b iv) also breaks TRS and MS, and has been explored extensively in the coherent control community for BPGE photocurrent generation~\cite{Hach1997,Atanasov1996,Rioux2011,Sun2010}, the individual symmetry breaking elements are less clear than for case iii, and the combined field does not exhibit chirality~\cite{Neufeld2025}, making it a more specific case for this tailored-light photocurrent spectroscopy technique. 

To further analyze these result, we performed \textit{ab-initio} simulations in similar conditions to the experiment, denoted by the solid lines in Fig.~\ref{Fig2}b. 
These agree remarkably well with our observations, and corroborate the selection-rule origin of the effects (we rule out other potential contributions such as electronic correlations in ref.~\cite{lesko2025}). 

Analytically, we can derive the consequences of TRS on induced BPGE currents by analyzing its impact on excited carrier distribution throughout the Brillouin zone (BZ), generated by the bichromatic field. 
Ultimately, the symmetry of the conduction (and valence) band's carrier \textit{k}-space population determines whether a bulk photogalvanic current will be generated. 
In this case, we recall that \textbf{$k$} and -\textbf{$k$} are connected by TRS (with a complex conjugation operation connecting the Bloch states, i.e. $\psi_\textbf{k}=\psi_{-\textbf{k}}^{*}$, where $\psi_\textbf{k}$ is the Bloch state at $\textbf{k}$), meaning that band curvatures and velocities at $\textbf{k}$ and $-\textbf{k}$ are exactly inverted from one another for a given band. 
Moreover, because graphene is both inversion and time-reversal symmetric (zero Berry curvature throughout the BZ) any arising photocurrent can only be a result of electronic occupation in excited bands and their potential asymmetric configuration in the BZ (i.e. an injection current). 
However, a laser field preserving TRS, applied in a material preserving TRS, must excite an equal amount of carriers at $\textbf{k}$ and $-\textbf{k}$, for any point $\textbf{k}$ in the BZ due to the degeneracy in the Bloch states being preserved by the laser driving. 
Critically, this means that TRS-preserving laser fields cannot generate BPGE photocurrents since the current generated by electrons occupying a given point in the BZ must be balanced by a reciprocal point.
This selection rule is quite striking, especially considering the fact that the driving laser electric field has a clear two-dimensional and spatially asymmetric configuration (see exemplary Lissajous in Fig.~\ref{Fig2}b ii), which \textit{a-priori} appears as if it should excite a photocurrent. 
The photocurrent measurements presented in Fig.~\ref{Fig2}a represent the first experimental observation of the TRS-induced selection rule for photocurrent generation by tailored light. 
Especially, it is the first observation of this selection rule in a light-matter system where other symmetries of the laser drive that could preclude photocurrents are absent (e.g. rotational or mirror planes that typically arise in monochromatic light).

Next, we show that the TRS induced BPGE photocurrent suppression selection rule can be employed through symmetry-breaking spectroscopy to probe TRS breaking phenomena. 
The logic of our proposal is as follows: Given an unknown sample where one would like to identify whether TRS is upheld or not (e.g. a magnet), we should irradiate it with the tailored $\mathbf{E}_{\omega-2\omega} (t)$ field while scanning \tr\ and \pr.
If no photocurrents are observed at the correct \pr\ for all \tr (\pr$=\pi/2$), TRS is preserved. Otherwise, it is broken, and the extent to which it is broken should correlate with the degree of intrinsic TRS breaking in the sample.
Remarkably, this approach does not require actually breaking TRS in the probe pulse, as is commonly done either in circular dichroism measurements or by applying external magnetic fields - the critical information is embedded at the parameter point where the two-color driving field respects TRS.
It is also inherently ultrafast time-resolved, and generates a background free signal for the symmetry-broken phases.

\begin{figure}[h!]
	\includegraphics[width=\linewidth]{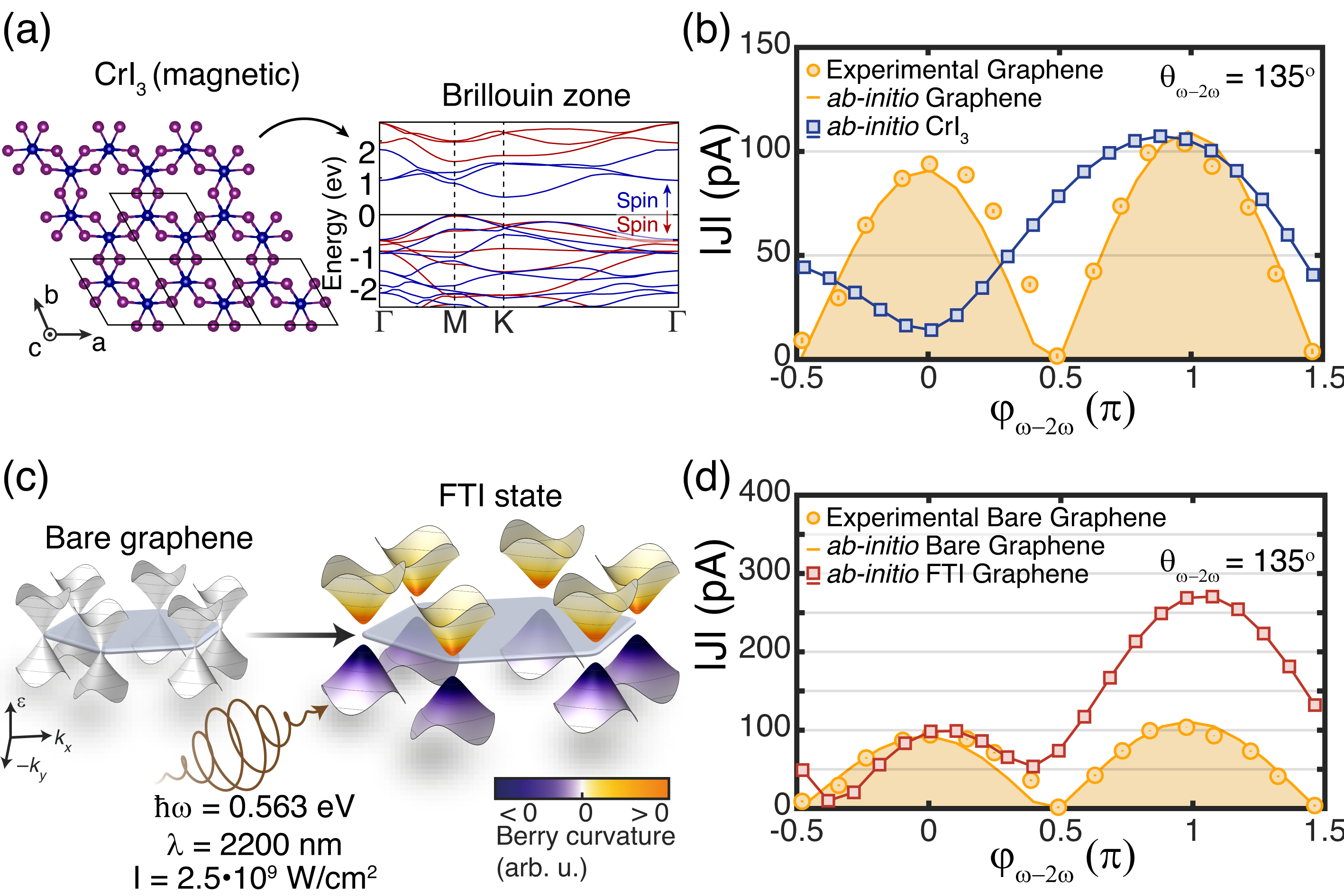}
    \caption{\textbf{Extension to time-reversal symmetry-broken systems.} 
    \textbf{a}, Real space depiction of the two-dimensional ferromagnet CrI$_3$, and its Brillouin zone. Blue and purple atoms represent chromium and iodine, respectively, while the blue and red bands in momentum space are for spin up/down, respectively. 
    \textbf{b}, \textit{ab-initio} TDDFT simulation of \pr-dependent photocurrents in Cr$_3$ at \tr$=135^\circ$ overlaid with both measured and simulated graphene photocurrents at identical laser parameters. See text for discussion.
    \textbf{c}, Momentum space depiction of optically dressed graphene with a circular field at 2200~nm, resulting in an FTI state (with 100~meV bandgap) with broken time-reversal symmetry, which we  probe by tailored light photocurrent spectroscopy.
    \textbf{d}, \textit{ab-initio} TDDFT simulation of \pr-dependent photocurrents in the FTI state at \tr$=135^\circ$ overlaid with both measured and simulated graphene photocurrents at identical laser parameters. Clearly, we can detect the TRS breaking of the FTI state.
    }
    \label{Fig3}
\end{figure}

We now test this concept in \textit{ab-initio} simulations (see Fig.~\ref{Fig3}).
First, we turn our attention to a two-dimensional magnet, CrI$_{3}$~\cite{Huang2017}.
From a material point of view, CrI$_{3}$ comprises a 2D hexagonal lattice similar to graphene, with the lattice structure and electronic band structure presented in Fig.~\ref{Fig3}a (see Methods section for technical details). 
Importantly, it is ferromagnetic in its ground state with nonzero magnetic moments on Cr $d$-states, which inherently breaks TRS. 
From a light-matter interaction perspective, these magnetic moments should cause deviations from the expected photocurrent suppression selection rule fundamentally due to the broken TRS symmetry.
Indeed, this deviation is present in the \pr-dependent photocurrents: In Figure~\ref{Fig3}b, we compare the \pr-dependent photocurrent generation in graphene (yellow, experimental and \textit{ab-initio}), to the predicted photocurrents from CrI$_{3}$ (blue, \textit{ab-initio}), for a \tr$=135^\circ$.
We observe a strong deviation from the typical curve observed in graphene: The minimal photocurrent response is predicted at \pr$=0$ instead of \pr$=\pi/2$, and it never reaches a zero value. 

Next, we investigate a TRS-broken system resulting from optical dressing, Fig.~\ref{Fig3}c.
By irradiating (i.e. dressing) the graphene sample with circularly-polarized mid-infrared light (at 2200~nm), we cause it to enter into a Floquet Chern insulating state~\cite{Oka2009,Lindner2011}, opening a topological gap.
Notably, due to the dressing field's rotational symmetry, it cannot induce a photocurrent in graphene on its own.
With these dressing conditions, the topological gap opening from the circular beam is small, but observable ($\sim 0.1$ eV for the highest field strength). 
Importantly, the $\omega-2\omega$ field in the symmetric configuration (Fig.~\ref{Fig1}c III, \tr$=135^\circ$, \pr$=\pi/2$) preserves TRS, and therefore cannot open a topological gap on its own~\cite{Neufeld2025}. 
By exploring the photocurrent response with respect to the $\omega-2\omega$ relative phase (at \tr$=135^\circ$, Fig.~\ref{Fig3}d) we observe a distinct deviation from the typical selection rule: Non-vanishing photocurrents arise for any choice of \pr, starkly different than the bare graphene case both from the \textit{ab-initio} simulation and experimentally measured. 
The impact of TRS breaking from the long-wavelength field is directly evident when observing the nonzero photocurrent at \pr$=\pi/2$ as well as the increased asymmetry of the \pr-dependent photocurrent. 
Furthermore, the minimum \pr-dependent photocurrent is slightly shifted to \pr$=0.45\pi$, reflecting that TRS is only marginally broken in the Floquet topological state. 
However, it proves that symmetry-breaking spectroscopy can be applied to probe driven quantum phases of matter. 
This suggests that such a spectroscopy scheme also encodes information about the Floquet topological gap size and the bands' Berry curvature that is imprinted onto the magnitude and sign of the photocurrent.

\begin{figure}[h!]
	\includegraphics[width=\linewidth]{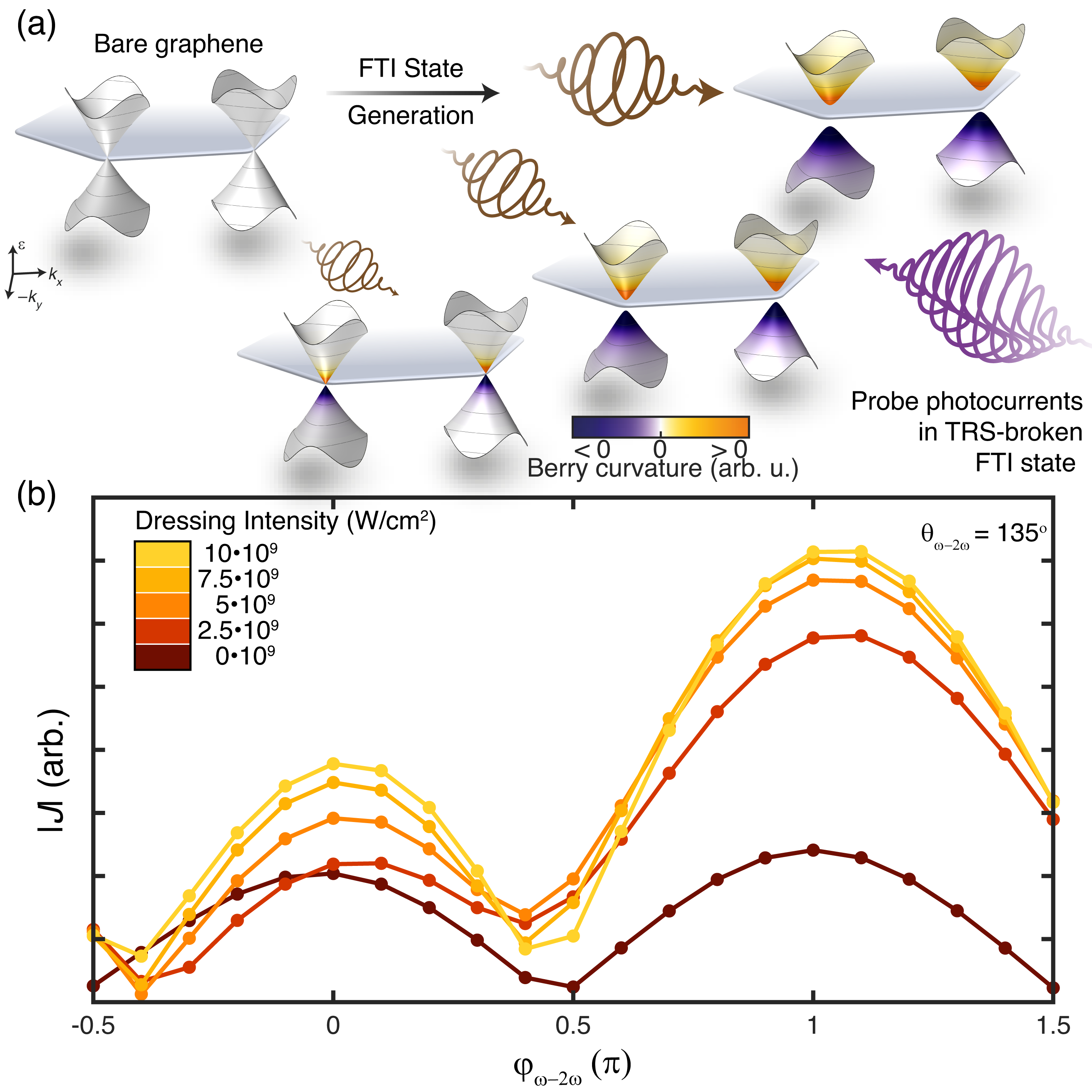}
    \caption{\textbf{Probing time-reversal symmetry in an FTI.} 
    \textbf{a}, Schematically, bare graphene is illuminated with a circular field with a central wavelength of 2200~nm. 
    As the dressing field increases intensity, the band gap increases and the Berry curvature spreads from the K/K' points (seen by the colors of the bands).
    \textbf{b}, Simulated ab-inito two-color photocurrents from dressed graphene, at a variety of dressing intensities (color bar for scale). Three key observations can be made. (i), Trivially, as the dressing field increases, more photocurrent is generated. (ii), Due to the broken time-reversal symmetry, the current minimum at \pr$= 0.5 \pi$ shifts to \pr$=0.45\pi$. (iii) Finally, the degree of time-reversal breaking can be seen from the decrease (and then reversal) of the dressing field intensity-dependent photocurrent at \pr$=-0.45\pi$. We observe this behavior because the degree of time-reversal symmetry breaking for the photocurrent-generating waveform is similar in magnitude but opposite in sign to the degree of time-reversal symmetry breaking of the FTI state.
    }
    \label{Fig4}
\end{figure}

We further explore the \pr-dependent photocurrent generation in an FTI by varying the intensity of the 2200~nm dressing pulse. 
As the intensity of the dressing field increases, it has a two-fold effect on the FTI state. First, the gap size at K/K' increases with the applied dressing field~\cite{Oka2009,Lindner2011}. Secondly, the Berry curvature of the bands spreads from K/K' as the dressing field increases, shown schematically in Fig.~\ref{Fig4}a.
Through \textit{ab-initio} simulations we predict \pr-dependent photocurrents for \tr$=135^\circ$ as a function of the dressing intensity (Fig.~\ref{Fig4}b).
Three key features emerge. The first is that the photocurrent magnitude increases for increasing dressing intensity, as intuitively expected.
The second is that the minimum photocurrent shifts away from the time-reversal symmetric waveform minimum of \pr$=\pi/2$.
Finally, the minimum photocurrent (at \pr$=-0.45\pi$) is non-zero for low dressing intensities, decreases as the dressing field increases, and increases again as the dressing field reaches its maximum.
This could be thought of as a measure of the degree of time-reversal symmetry breaking of the FTI state.
When the FTI's state's time-reversal symmetry is weakly broken, a biharmonic field with the opposite chirality can effectively restore the symmetry of the system (albeit, at a different \pr~ than the TRS maintaining system).
While the total combined (dress and optical two-color probe) optical waveform is exceedingly complex, this \pr-dependent photocurrent could represent a direct probe of both the magnitude and sign of the degree of time-reversal symmetry breaking in light dressed states, correlating to the gap size and Berry curvature sign of the FTI state. 

Our simulations, complemented by experiential observations in graphene, confirm the ability to probe TRS-broken phases using tailored TRS-respecting light probes (absent of circularly-polarized components). 
Most importantly, the signal is effectively background free and potentially ultrafast time-resolved. 
To our knowledge, this marks the first approach that provides such capabilities and should open up new opportunities for exploring ultrafast magnetism~\cite{Walowski2016}, steady-state Floquet phases driven by monochromatic~\cite{McIver2019,Liu2025} or tailored light~\cite{Neufeld2019B}, phonon driven magnetism~\cite{Nova2016,Luo2023}, as well as by other magnetic mechanisms~\cite{Song2025}, and finally chirality-induced-spin-selectivity~\cite{Bloom2024}.

\vspace{\baselineskip}
\noindent \textbf{Discussion and outlook}

To summarize, we experimentally and theoretically explored tailored-light-driven bulk photogalvanic currents in graphene and predicted their response in other hexagonal 2D systems. 
We showed that for a unique case, where the tailored field comprises bichromatic linearly polarized $\omega-2\omega$ components, one can engineer time-reversal symmetry (TRS) and mirror symmetry respecting/breaking laser pulses by scanning both the polarization angle (\tr) and two-color phase (\pr). This showcases the biharmonic field's ability to probe phases of matter without explicitly breaking their symmetry, as commonly done with monochromatic fields.
For the first time, we measured this tailored-light photocurrent response in graphene with respect to \tr~and \pr, allowing us to demonstrate TRS-induced photocurrent suppression selection rules. 
This is the first observation of such selection rules, and is unique in the ability to isolate the effect of TRS from other symmetries in the laser field that typically suppress photocurrent responses (e.g. mirror, inversion, and rotational symmetries). 
We analyzed this result both analytically and with state-of-the-art \textit{ab-initio} simulations, all of which corroborated our interpretation of the physical mechanism.
Utilizing tailored light photocurrent spectroscopy we performed calculations in more complex quantum systems, showing that the TRS-induced selection rule is broken in TRS-broken systems such as a 2D magnet and a Floquet Chern insulator (correlating with the topological gap size, and the degree of time-reversal symmetry breaking). 
Thus, our work predicts clear signatures of TRS-broken phases in tailored light photocurrent spectroscopy, and to our knowledge is the first technique that permits probing such phases without circularly polarized fields and/or external magnetic fields, which can themselves alter the state being explored. Thus, our work should allow enhanced accuracy spectroscopy of ultrafast magnetism\cite{Walowski2016,Siegrist2019,Neufeld2023} and phase transitions.  

Our work should also impact other ultrafast spectroscopies of various physical and chemical systems such as chirality~\cite{Habibovi2024} and chirality-induced spin selectivity ~\cite{CISS}, where novel methods are needed. 
Looking forward, we believe these results will bridge the fields of highly-nonlinear optics in solids exploring tailored light responses such as high harmonic generation~\cite{Dudovich2006,Ayuso2019}, and nonlinear photocurrent excitation~\cite{Pettine2023}, allowing the two effects to be analyzed on an equal footing~\cite{Neufeld2021,Neufeld2025}.


\vspace{\baselineskip}
\noindent \textbf{Acknowledgments:}
The authors thank Dr. Lu Wang for her comments on the manuscript.
ON acknowledges the scientific support of Prof. Dr. Angel Rubio. DMBL acknowledges support from the Emerging Talents Initiative at FAU. 
This work has been funded by the Deutsche Forschungsgemeinschaft (SFB 953 ‘Synthetic Carbon Allotropes’, 182849149), the PETACom project financed by Future and Emerging Technologies Open H2020 program, ERC Grant AccelOnChip (884217), and the Gordon and Betty Moore Foundation (GBMF11473).

\vspace{\baselineskip}
\noindent \textbf{Corresponding authors:}
Correspondence to 
\href{mailto:daniel.lesko@fau.de}{Daniel Lesko},
\href{mailto:peter.hommelhoff@fau.de}{Peter Hommelhoff}, and
\href{mailto:ofern@technion.ac.il}{Ofer Neufeld}.

\vspace{\baselineskip}
\noindent \textbf{Methods}

\noindent \textbf{Experimental methods}

For producing the biharmonic tailored fields, we frequency double an Erbium fiber frequency comb~\cite{Lesko2021} at 80~MHz in a 1~mm BiBO crystal producing 213~fs and 110~fs pulses for the fundamental (1550~nm) and second harmonic (775~nm) respectively, identical to ref.~\cite{lesko2025}. 
The delay between the two fields (\pr) is controlled by a calcite inline interferometer~\cite{Brida2012}, while the angle of the second harmonic is controlled by a two-color half waveplate (Newlight Photonics, \tr), shown schematically in Fig.~\ref{Fig1}b.
The biharmonic tailored fields are then focused with an off axis parabolic mirror to 0.27 and 0.2~V/nm, for the fundamental and second harmonic respectively, on monolayer epitaxially grown graphene on SiC.
The graphene on SiC is placed within vacuum chamber at 10$^{-8}$ hPa and at room temperature. 
Finally, \pr-dependent photocurrents (green arrow) are measured using gold electrodes in the $x$-direction using a lock-in amplifier referenced to the \pr.
Further experimental details are provided in ref.~\cite{lesko2025}.

\noindent \textbf{TDDFT simulations}

This section describes technical details of \textit{ab-initio} simulations presented in the main text. All calculations were performed with the open access code Octopus~\cite{TancogneDejean2020}. The ground state of graphene and CrI$_{3}$ were obtained at the hexagonally-symmetric lattice with lattice parameters 4.65 and 7 Bohr, respectively, where otherwise symmetries over Bloch states were not imposed. $\Gamma$-centered k-grids were employed of sizes 120$\times$120 and 15$\times$15 were employed in graphene and CrI$_{3}$, respectively. We employed grid spacings of 0.38 and 0.39 Bohr, and a z-axis of size 60 and 50 Bohr, for graphene and CrI$_{3}$, respectively. Time-dependent simulations were initiated from the ground-state electron configuration at $t=0$. For CrI$_{3}$ we employed spin-DFT, leading to a magnetic ground state with a magnetization of ~3.2 Bohr on each Cr atom and a semiconducting phase.

We solved the time dependent Kohn-Sham (KS) equations of motion on a real-space grid in the length gauge:
\begin{widetext}
\begin{equation}
\label{eq:kseom}
	i\partial_{t}\psi_{n,\textbf{k}}(\mathbf{r},t) =	
        (\frac{1}{2}(-i\bm{\nabla} + \textbf{A}(t)/c)^2 + v_\mathrm{KS}(\mathbf{r},t))\psi_{n,\textbf{k}}(\mathbf{r},t),
\end{equation}
\end{widetext}
where $\psi_{n,\textbf{k}}(\mathbf{r},t)$ is the KS (KS) state at band $n$ and \textit{k}-point $\textbf{k}$, and $v_\mathrm{KS}$ is the KS potential comprising a classical Hartree term, the interactions of electrons with nuclei and deeper core states (incorporating norm-conserving pseudopotentials~\cite{Hartwigsen1998}), and the exchange-correlation (XC) potential. We employed the adiabatic local density approximation (aLDA) for the XC functional throughout. Periodic boundaries were employed in the monolayer plane while the \textit{z}-axis was treated with finite boundary conditions. A complex absorber of width 12 Bohr was added during propagation (similar to the approach in ref.~\cite{Neufeld2021}). In eq.~\ref{eq:kseom}, $\mathbf{A}(t)$ is the applied vector potential within the dipole approximation:
\begin{widetext}
\begin{equation}
\label{eq:vector}
\mathbf{A}(t)=f(t)\frac{cE_{0}}{\omega}(\cos(\omega t+\varphi_{\omega-2\omega})\hat{\mathbf{y}} + \frac{\Delta}{2}(\cos(\theta)\cos(2\omega t)\hat{\mathbf{x}}+\sin(\theta)\sin(2\omega t)\hat{\mathbf{y}}))
\end{equation}
\end{widetext}
where $c$ is the speed of light, $E_{0}$ is the electric field amplitude (taken at the experimental values), $\Delta=0.75$ is the field amplitude ratio, $\varphi_{\omega-2\omega}$ the two-color phase, $\epsilon$ the ellipticity of the $2\omega$ field, $\omega$ the carrier frequency (taken at the experimental value corresponding to 1550 nm light), and $f(t)$ a temporal envelope taken as a 'super-sine' form~\cite{Neufeld2019C}: 
\begin{equation}
\label{eq:env}
f(t) = \sin{\left(\pi \frac{t}{T_\mathrm{p}}\right)}^{\left(\frac{|\pi(\frac{t}{T_\mathrm{p}}-\frac{1}{2})|}{\sigma}\right)}
\end{equation}
where $\sigma=0.75$, $T_\mathrm{p}$ is the laser pulse duration chosen as $T_\mathrm{p}=8T$, and $T=\SI{5.17}{\femto\s}$ is a single cycle of the $\omega$ carrier frequency. The angle $\theta$ is the relative angle between the two carrier wave polarizations, which are taken to be linearly-polarized at all configurations. 

The KS equations were propagated, from which we obtained the time-dependent KS states. The total time-dependent light-driven photocurrent was calculated by: 
\begin{widetext}
\begin{equation}
\label{eq:curr}
\textbf{J}(t) = \sum_{n,k}w_{k}\int [\psi^{*}_{n,\textbf{k}}(\mathbf{r},t)(\frac{1}{2}(-i\bm{\nabla} + \textbf{A}(t)/c) - i[V_\mathrm{ion},\textbf{r}])\psi_{n,\textbf{k}}(\mathbf{r},t)]\mathrm{d}\textbf{r}  + c.c.,
\end{equation}
\end{widetext}
where $V_{ion}$ is the non-local part of $v_{KS}$ (due to the pseudopotentials), $w_{k}$ is the \textit{k}-point weight, and the sum is performed over occupied KS states. 
From \textbf{J}$(t)$ the injection current was evaluated by averaging over a single cycle of the fundamental frequency after the driving laser pulse has ended (at $t=t_f$): \textbf{j}$_{photo}=\int_{t_f}^{t_f+T} \textbf{J}(t){d}t$.

In the case of the Chern insulating phase, the same procedures were employed as in the bare graphene simulations, except that an additional laser pulse centered at 2200~nm with a circular polarization the \textit{xy}-plane was incorporated, and with the same temporal envelope.

\end{document}